\documentclass[journal]{IEEEtran}

\usepackage[cmex10]{amsmath}
\usepackage{amssymb}
\usepackage{textcomp}
\usepackage[pdftex]{graphicx}
\usepackage{cite}
\usepackage{hyperref}
\usepackage{xcolor}
\usepackage[caption=false, font=footnotesize]{subfig}

\usepackage{mathtools}  

\begin{document}

\title{Concatenated Reed-Solomon and Polarization-Adjusted Convolutional (PAC) Codes}

\author{Mohsen~Moradi\textsuperscript{\href{https://orcid.org/0000-0001-7026-0682}{\includegraphics[scale=0.06]{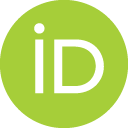}}},
Amir~Mozammel\textsuperscript{\href{https://orcid.org/0000-0003-3474-9530}{ \includegraphics[scale=0.06]{Figures/ORCIDiD_icon128x128}}}
\thanks{The authors are with the Department of Electrical-Electronics Engineering, Bilkent University, Ankara TR-06800, Turkey (e-mail: moradi@ee.bilkent.edu.tr, a.mozammel@ee.bilkent.edu.tr).}%
}

\maketitle

\begin{abstract}
Two concatenated coding schemes incorporating algebraic Reed-Solomon (RS) codes and polarization-adjusted convolutional (PAC) codes are proposed. 
Simulation results show that at a bit error rate of $10^{-5}$, a concatenated scheme using RS and PAC codes has more than $0.25$ dB coding gain over the NASA standard concatenation scheme, which uses RS and convolutional codes.
\end{abstract}


\section{Introduction}
\IEEEPARstart{R}{eed-Solomon} (RS) codes are a class of algebraic block-based error-correcting codes with a wide range of applications in digital communications and storage \cite{reed1960polynomial}. 
In this paper, by benefiting from the RS decoder's burst-error-correction capability, we introduce two concatenation schemes to improve the error-correction performance of polarization-adjusted convolutional (PAC) codes \cite{arikan2019sequential}.

\begin{figure}[t] 
\centering
	\includegraphics [width = \columnwidth]{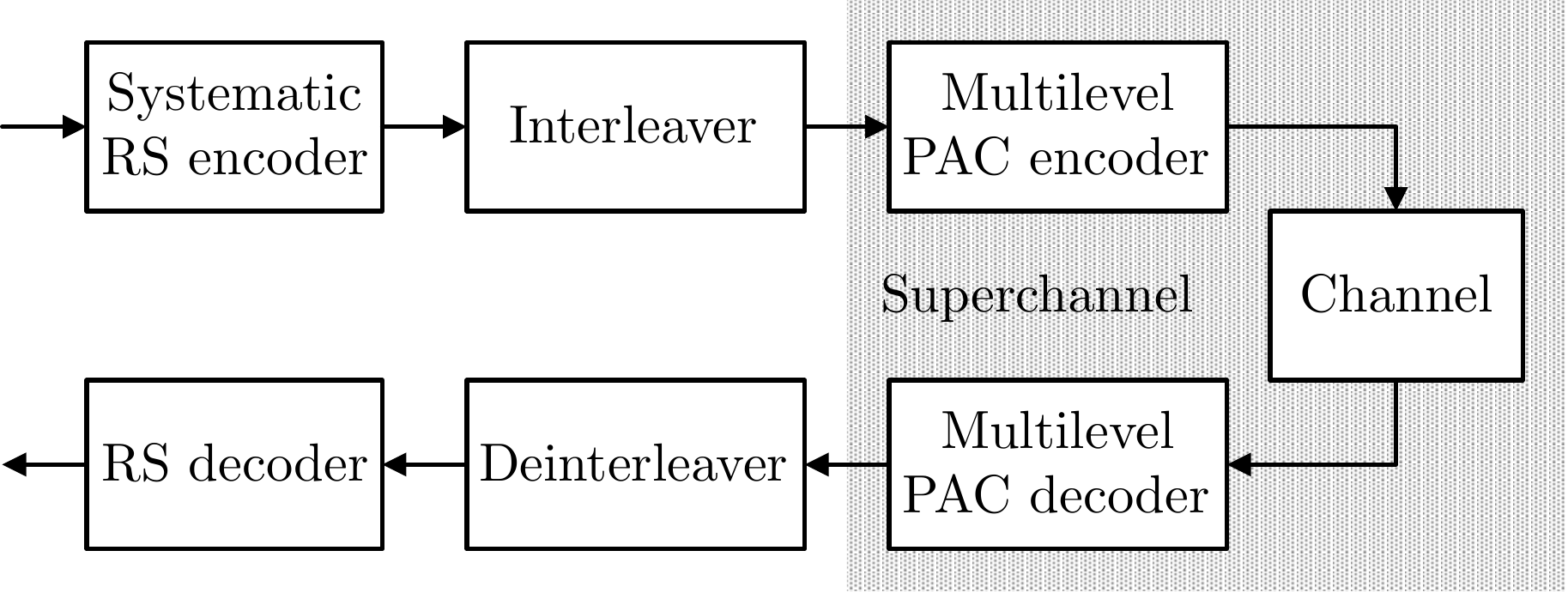}
	\caption{Block diagram of concatenated coding scheme.} 
	\label{flowchart: concatenated}
\end{figure}

Fig. \ref{flowchart: concatenated} demonstrates the general concatenation scheme of RS and PAC (RS-PAC) codes.
Using multilevel PAC code as an inner code makes the RS code see a superchannel (consisting of a multilevel PAC encoder, physical channel, and a multilevel PAC decoder). 
A PAC code of block length $N$ may introduce a burst error of size up to $N$, and parameters of the outer RS code should be defined such that it can correct these burst errors. 
Also, using an interleaver helps to scatter the error bursts of lengths up to $N$ of the superchannel between different RS codes and further improves the error-correction capability of the concatenated RS-PAC codes. 

Devised by Forney \cite{forney1965concatenated}, concatenated codes mainly use an RS code as the outer code to handle the burst errors.
Using a convolutional code (CC) as an inner code under Viterbi decoding introduces a short burst of errors of size up to the encoder memory $m$.
Constructing the RS code over the field GF$(2^s)$ with symbol length $s$ such that $s > m$ helps the concatenated code handle the burst bit errors which could not be corrected by the inner CC.

Employing a block code like a Reed-Muller (RM) code with soft-decision decoding \cite{morelos1999constructions} or a convolutional code under sequential decoder as an inner code \cite{falconer1969hybrid} makes the superchannel introduce some scattered random errors. 
For this reason, inner code generally has a short block length, and the resulting concatenated code may have worse error-correction performance in comparison with an RS-CC under Viterbi decoding.

Falconer \cite{falconer1969hybrid} suggested using CC of length $N = I\times n^{'}$ under sequential decoding as the inner code.
As this inner code introduces scattered errors similar to the block codes, Falconer suggested constructing RS code over GF$(2^N)$ or alternatively using $I$ parallel RS codes over GF$(2^{n{'}})$.
This requires a large number of parallel RS codes and may result in a worse tradeoff between the complexity and error-correction performance compared to using a CC under Viterbi decoding.

PAC codes under sequential decoding have variable computational complexity.
However, the average computational complexity is low \cite{moradi2020metric}, which makes the PAC codes suitable as an inner code in the multilevel concatenated scheme. Performance results show that our introduced concatenated RS-PAC codes have a better error-correction performance than an RS-CC with much better average computational complexity.
Simulation results also show that an RS-PAC code has an error-correction performance comparable to an RS-RM code with much better computational complexity.

Throughout this paper, PAC, RM, and CCs are over the binary Galois field GF$(2)$, and RS codes are over GF$(2^8)$. 
We use boldface notation for the matrices and vectors. For a vector $\mathbf{x} = (x_1, x_2, ..., x_N)$, $\mathbf{x}^i$ denotes subvector $(x_1, x_2, ..., x_i)$.
For any subset of indices $\mathcal{A} \subset \{1, 2, ..., N\}$, $\mathbf{x}_{\mathcal{A}}$ represents subvector $(x_i : i\in \mathcal{A})$, and $\mathcal{A}^c$ denotes the complement of set $\mathcal{A}$.
For a matrix $\mathbf{G}$, $\mathbf{G}_{\mathcal{A},\mathcal{B}}$ denotes a submatrix of $\mathbf{G}$ that rows are selected by set $\mathcal{A}$, and columns are selected by set $\mathcal{B}$.

The remainder of this paper is organized as follows. 
Section \ref{sec: PAC scheme} reviews blocks and parameters of PAC codes.
In Section \ref{sec: RS codes}, encoding and decoding of RS codes are briefly reviewed.
Section \ref{sec: RS-PAC Concatenated} proposes two concatenated RS-PAC codes and provides simulation results and comparisons.
Finally, Section \ref{sec: Conclusion} concludes the paper.

\section{PAC Codes}\label{sec: PAC scheme}

\begin{figure}[t] 
\centering
	\includegraphics [width = \columnwidth]{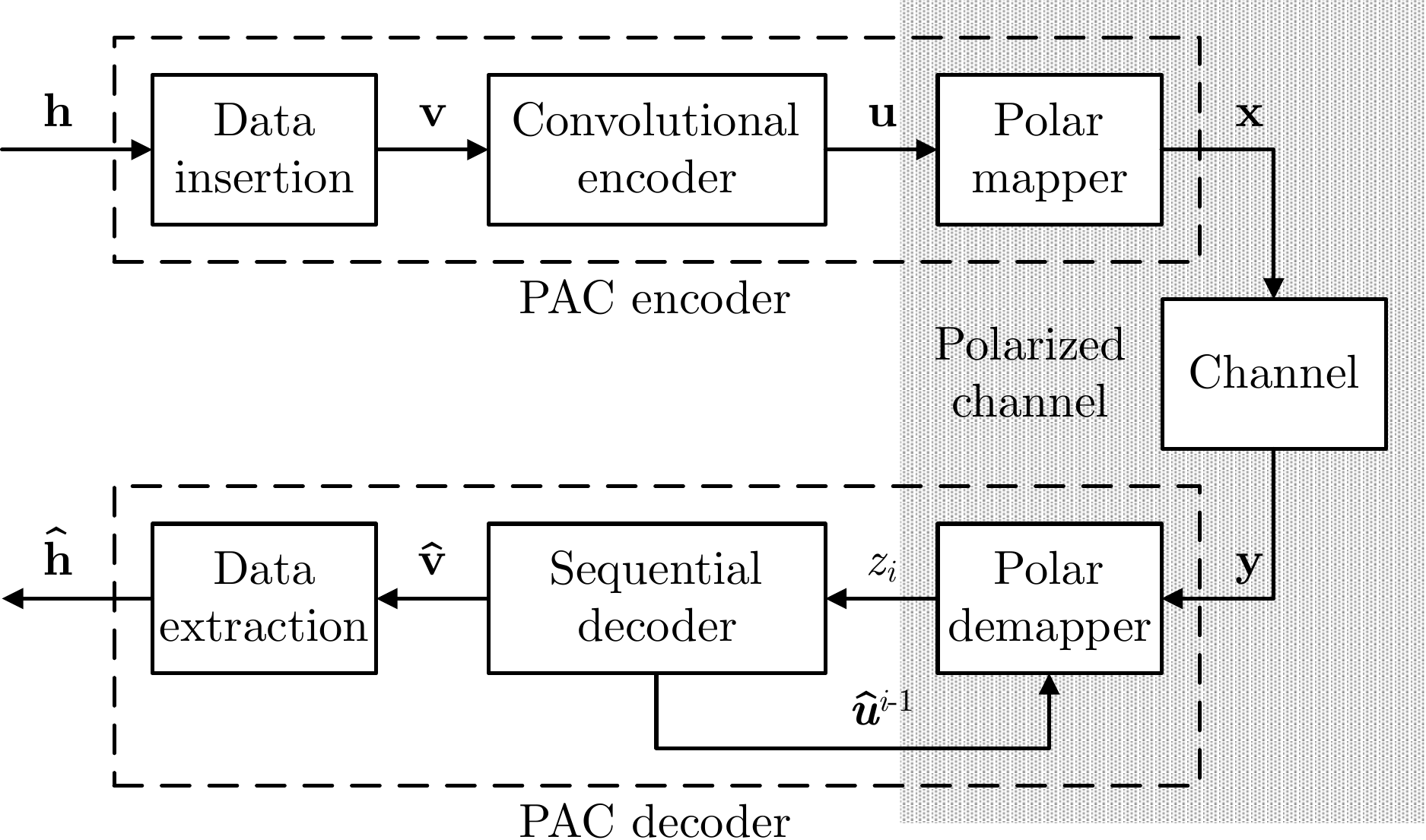}
	\caption{Block diagram of PAC coding scheme.} 
	\label{fig: PAC flowchart}
\end{figure}

Fig. \ref{fig: PAC flowchart} shows a block diagram of a $(N,K,\mathcal{A}, \mathbf{c})$ PAC coding scheme, where $N$ is the code length, $K$ is the data word length, $\mathcal{A}$ is the data indexing set (a.k.a. rate profile), and $\mathbf{c}$ is the connection polynomial of convolutional code.
In this paper, we use $\mathbf{c} = 3211$ (in octal form) \cite{moradi2020PAC}.
$\mathbf{h}=(h_{1},\ldots,h_{K})$ is the source word generated uniformly at random over all possible source words of length $K$ in a binary field GF$(2)$.
The data insertion block maps these $K$ bits into a data carrier vector $\mathbf{v}$ in accordance with the data set $\mathcal{A}$, thus inducing a code rate of $R=K/N$.
In this paper, for PAC$(128, 64)$ codes, an RM rate profile \cite{arikan2019sequential} is used.
For PAC$(64, 32)$ and PAC$(64, 40)$ codes we obtain the rate profiles according to the method introduced in \cite{moradi2021monte} at $5$ and $6$ dB signal-to-noise ratio (SNR) values, respectively.  
In a non-systematic PAC encoder, after $\mathbf{v}$ is obtained by $\mathbf{v}_\mathcal{A} = \mathbf{h_j}$ and $\mathbf{v}_{\mathcal{A}^c} = 0$, it is sent to the convolutional encoder and encoded as $\mathbf{u} = \mathbf{v}\mathbf{T}$, where $\mathbf{T}$ is an upper-triangular Toeplitz matrix with first row obtained by the connection polynomial $\mathbf{c}$.
Then $\mathbf{u}$ is transformed to $\mathbf{x}$ with standard polar transformation $\mathbf{F}^{\otimes n}$, where $\textbf{F}^{\otimes n}$ is the $n$th kroneker power of $\textbf{F} = \begin{bsmallmatrix} 1 & 0\\ 1 & 1 \end{bsmallmatrix}$ with $n = \log_n N$.

In this paper, we use a systematic PAC encoder \cite{arikan2020systematic}, in which the data word $\mathbf{h_j}$ is encoded to $\mathbf{x}$ as
\begin{equation}
    \mathbf{x}_\mathcal{A} = \mathbf{h},
    ~~~~~~~
    \mathbf{x}_{\mathcal{A}^{c}} = \mathbf{h}
    \left(\mathbf{G}_{\mathcal{A},\mathcal{A}} \right)^{-1}
    \mathbf{G}_{\mathcal{A},\mathcal{A}^c},
\end{equation}
where $\mathbf{G} = \mathbf{T}\mathbf{F}^{\otimes n}$.
After the PAC encoding, $\mathbf{x}$ is sent through the channel.
Polar demapper receives the channel output $\mathbf{y}$ and the previously decoded bits and calculates the log-likelihood ratio (LLR) value of the current bit $z_i$.
Finally, the sequential decoder outputs an estimate of the carrier word $\hat{\mathbf{v}}$, from which the $K$-bits data can be extracted according to $\mathcal{A}$.

By adopting the notation used in \cite{moradi2020metric}, we use the partial path metric for the first $i$ branches as
\begin{equation} \label{partialmetric}
    \Gamma(\mathbf{u}^i;\mathbf{y}^N) = \log_2 \left( \frac{P(\mathbf{y}^N | \mathbf{u}^{i})}{P(\mathbf{y}^N)}\right) - 
    \sum_{j=1}^{i}E_0(1, W_N^{(j)}),
\end{equation}
where $\mathbf{y}^N$ is the channel output, $\mathbf{u}^i$ is the partial convolution output, and $E_0(1, W_N^{(j)})$ is the $j$th bit-channel cutoff rate \cite{moradi2020metric}.

\section{Reed-Solomon Codes}\label{sec: RS codes}
RS codes are nonbinary linear block error-correcting codes and are a subset of nonbinary BCH codes \cite{reed1960polynomial, bose1960class}.
A $(n, k, d_{min})$ RS code over GF$(2^s)$ guarantees to correct up to $t = \lfloor\dfrac{d_{min}-1}{2} \rfloor$ symbol errors, where $n$ is the RS code block length, $k$ is the data length, and $d_{min}$ is the minimum Hamming distance of the code.
A symbol that can be corrected by RS codes may have $1, 2, \cdots, s$ bit errors. 
For this reason, RS codes are pretty suited to correct burst errors (contiguous sequence of bits in error). 
As an example, a $(255, 223, 33)$ RS code over GF$(2^8)$ can correct up to 16 symbol errors (up to $16 \times 8$ bits). 
RS codes are maximum distance separable (MDS) codes; meaning that the $d_{min}$ of an RS code is the largest possible minimum distance for a given $n$ and $k$ ($d_{min} = n - k +1$).
When the number of symbol errors of a received data is less than $t$, an RS code's algebraic decoder will always correctly decode the received vector.
However, if the number of symbol errors exceeds $t$, the decoder may declare a decoding failure based on the distribution and number of errors. Otherwise, the decoder wrongly decodes to another valid codeword. 

In this paper, RS codes are defined over GF$(2^8)$, and each code symbol is one byte.
In our proposed concatenated RS-PAC code, we use a systematic encoder of RS code which takes $k$ data symbols ($8k$ bits) as input and appends $n-k$ parity symbols to make an $n$ symbol ($8n$ bits) codeword. 
We use primitive polynomial $\pi(x) = 1 + x^2 + x^3 + x^4 + x^8$ to represent the GF$(2^8)$. 
Let $\alpha$ be a primitive element in GF$(2^8)$ (order of element $\alpha$ is 255).
For a $(n, k, d_{min})$ RS code that is capable of correcting $t$ symbol errors, the code generator polynomial is
\begin{equation}
    g(x) = (x-\alpha)(x-\alpha^2) \cdots (x-\alpha^{2t}).
\end{equation}
The systematic encoding of RS code is as
\begin{equation}
    h(x) = m(x)x^{n-k} + P(x),
\end{equation}
where message polynomial
\begin{equation*}
    m(x) = m_0 + m_1x + \cdots m_{k-1}x^{k-1}
\end{equation*}
corresponds to the message $\mathbf{m} = (m_0, m_1, \cdots, m_{k-1})$, codeword polynomial
\begin{equation*}
    h(x) = h_0 + h_1x + \cdots h_{n-1}x^{n-1}
\end{equation*}
corresponds to the codeword $\mathbf{h} = (h_0, h_1, \cdots, h_{n-1})$, and the parity polynomial $P(x)$ is the remainder of polynomial $m(x)x^{n-k}$ after division by polynomial $g(x)$.
Since for a generator polynomial $g(x)$,
\begin{equation}
    g(\alpha ) = g(\alpha^2 ) = \cdots = g(\alpha^{2t} ) = 0,
\end{equation}
for a codeword polynomial $h(x)$ we have
\begin{equation}
    h(\alpha ) = h(\alpha^2 ) = \cdots = h(\alpha^{2t} ) = 0,
\end{equation}
and all codeword polynomials are divisible by $g(x)$. 

\begin{figure*}[t] 
\centering
	\includegraphics [width = 0.7\textwidth]{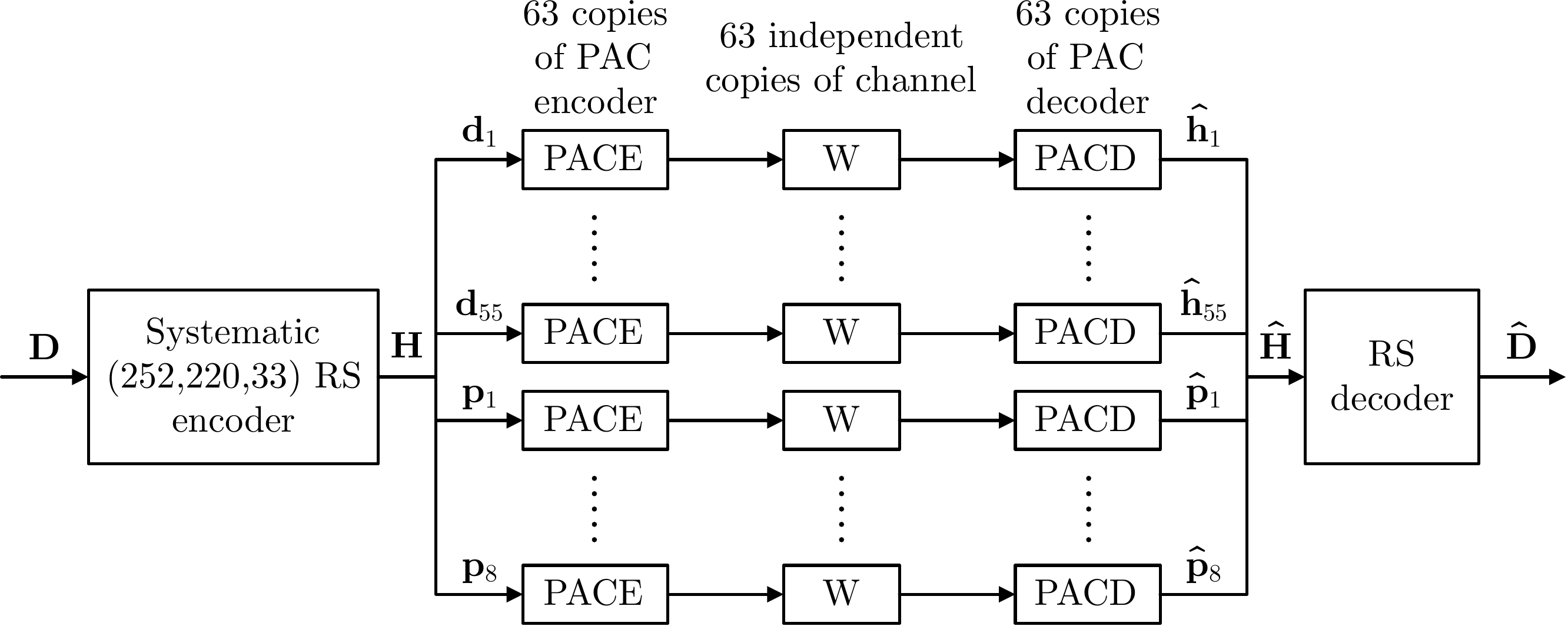}
	\caption{Block diagram of RS-PAC coding scheme.} 
	\label{fig: RS_PAC flowchart}
\end{figure*}

After receiving a channel output $r(x)$, an RS decoder attempts to correct up to $t$ symbol errors by identifying both the error locations and error values.
First step of decoding is to calculate $2t$ syndroms of the received polynomial $r(x)$ as
$S_j = r(\alpha^j)$ for $j = 1, 2, \cdots, 2t$.
Assume that the received channel output has $\delta$ symbol errors. 
The second step of decoding is to obtain the error locator polynomial $\Lambda(x)$ of degree $\delta$ which $\delta$ roots of $\Lambda(x)$ are the reciprocals of the error locations. 
Obtaining $\Lambda(x)$ can be done using the Euclidean or Berlekamp-Massey algorithm \cite{sugiyama1975method, berlekamp2015algebraic}. 
The Euclidean algorithm is easier to implement; however, the Berlekamp-Massey algorithm is more efficient for both hardware and software implementations. Berlekamp–Massey algorithm's computational complexity is of $\mathcal{O}(\delta^2)$ order. 

After obtaining $\Lambda(x)$, the decoder's job is to find its roots.
One inefficient way to determine the roots of $\Lambda(x)$ is to examine every element of the finite field to determine whether it is a root of the error locator polynomial.
Chien search is an efficient way to find the roots $X_i$ of $\Lambda(x)$ \cite{chien1964cyclic}.
If Chien search results in less than $\delta$ distinct roots (degree of $\Lambda(x)$), the decoding algorithm can declare a decoder failure (roots may be repeated or in an extension field of GF$(2^8)$). 

As RS code is a nonbinary code, in addition to finding the error locations $X_i^{-1}$, the decoder should determine the error values as well.
According to Forney's algorithm \cite{forney1965decoding}, error polynomial $e(x)$ is computed as
\begin{equation}
    e(x) = -\dfrac{\Omega(x)}{\Lambda^{'}(x)},
\end{equation}
where $\Omega(x) = S(x)\Lambda(x)~~(\text{mod}~x^{2t})$ is the the error-evaluator polynomial and $\Lambda^{'}(x)$ is the formal derivative of $\Lambda(x)$.
For the $i$th error location $X_i^{-1}$, $e_i = e(X_i^{-1})$ is the $i$th error value.
Finally, the recovered codeword polynomial is $\hat{h}(x) = r(x) + e(x)$.

\section{RS-PAC Concatenated Codes}\label{sec: RS-PAC Concatenated}
A single-level concatenated coding scheme usually employs a nonbinary code such as RS code as an outer code and a binary code such as CC as an inner code. 
For a CC of memory size $m$ as the inner code, a single incorrect decoding decision might give rise to a burst decoding error of length $m$. 
Benefiting from an RS code over GF$(2^m)$, a burst of $m$ bit errors introduced by the inner code is interpreted as one symbol error by the outer RS code.
An RS code capable of correcting $t$ symbol errors can correct up to $t$ of these burst errors. 
This concatenation results in a powerful code with excellent error-correction performance \cite{lin2001error}.
In a PAC code, because of the polarization effect, the CC sees a channel with a memory of $N$, and thus a wrongly decoded bit may result in a burst error of size up to $N$ bits. 

\subsection{RS-PAC Concatenated Codes without Interleaver}
This section constructs a  multi-level concatenated RS and PAC (RS-PAC) code as illustrated in Fig. \ref{fig: RS_PAC flowchart}.
Assume a $(252, 220, 33)$ RS code over GF$(2^8)$ with a systematic encoder. 
An input data $\mathbf{D} = \left( \mathbf{d}_1, \mathbf{d}_2,\cdots, \mathbf{d}_{55} \right)$ of length $220$ symbols is the input of both RS encoder and 55 parallel PAC$(64, 32)$ encoders, where vector $\mathbf{d}_i$ consists of 4 symbols (32 bits).
The overall code rate of this coding scheme is $R = (220/252)\times(32/ 64) \approx 0.44$.
The RS encoder calculates a parity $\mathbf{P} = \left( \mathbf{p}_1, \mathbf{p}_2, \cdots, \mathbf{p}_8 \right)$ of length 32 symbols which is the input of 8 other PAC encoders. 
Each vector $\mathbf{p}_i$ consists of 4 symbols (32 bits).
The whole RS systematic encoder output is 
\begin{equation*}
    \mathbf{H} = 
    \left(
    \mathbf{h}_1, \mathbf{h}_2,\cdots, \mathbf{h}_{63} \right)
    =
    \left( \mathbf{d}_1, \mathbf{d}_2,\cdots, \mathbf{d}_{55}, 
    \mathbf{p}_1, \mathbf{p}_2, \cdots, \mathbf{p}_8
    \right).
\end{equation*}
Overall, each of 63 parallel PAC encoders receives vector $\mathbf{h}_i$ of length 4 symbols (32 bits) for $i$ from 1 to 63.
The output of each PAC encoder is sent through 63 copies of the channel, and the channel outputs are decoded with the corresponding PAC decoder to obtain an estimate $\hat{\mathbf{h}}_i$ corresponding to $\mathbf{h}_i$. 
The output of 63 parallel PAC decoder is denoted by vector
\begin{equation*}
    \mathbf{\hat{H}} = 
    (
    \mathbf{\hat{h}}_1, \mathbf{\hat{h}}_2,\cdots, \mathbf{\hat{h}}_{63} 
    ),
\end{equation*}
where each $\mathbf{\hat{h}}_i$ has 4 symbols (32 bits), and $\mathbf{\hat{H}}$ has a length of 252 symbols. 
Finally, the RS decoder receives vector $\mathbf{\hat{H}}$ and outputs the estimate data 
\begin{equation*}
    \mathbf{\hat{D}} = 
    (
    \mathbf{\hat{d}}_1, \mathbf{\hat{d}}_2,\cdots, \mathbf{\hat{d}}_{55} ),
\end{equation*}
where $\mathbf{\hat{D}}$ has a length of 220 symbols.

Note that the $(252, 220, 33)$ RS code can correctly decode up to 16 symbol errors ($4\times 32$ bits). 
In the case that RS decoder declares a decoding failure, RS-PAC concatenated code outputs the first 220 symbols of RS decoder input 
$(\mathbf{\hat{h}}_1, \mathbf{\hat{h}}_2,\cdots, \mathbf{\hat{h}}_{55})$.
Alternatively, this RS-PAC coding scheme can be constructed using a $(240, 208, 33)$ RS code as the outer code, and PAC$(128, 64)$ or PAC$(256, 128)$ codes as the inner codes. 
The former uses 30 parallel PAC$(128, 64)$ codes, while the latter uses 15 parallel PAC$(256, 128)$ codes.

\begin{figure}[t] 
\centering
	\includegraphics [width = 0.95\columnwidth]{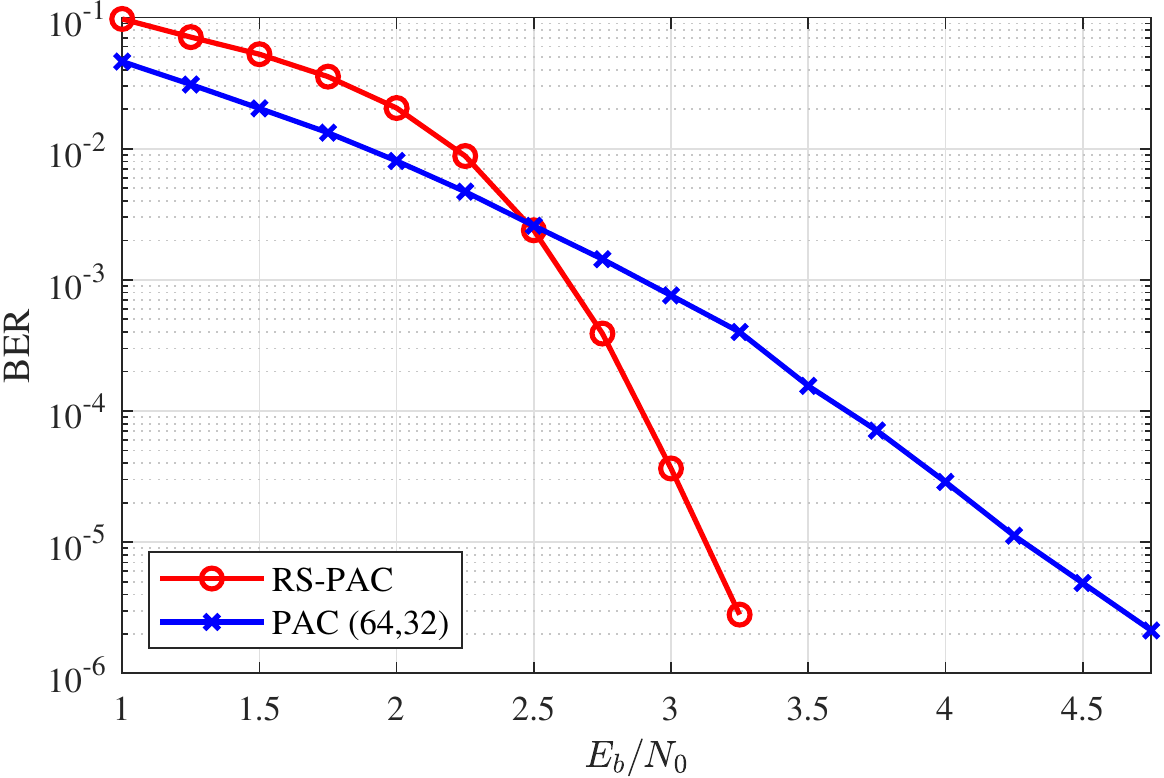}
	\caption{BER performance of RS-PAC concatenated code of scheme 1 v. PAC$(64, 32)$ code.} 
	\label{fig: BER}
\end{figure}

Fig. \ref{fig: BER} plots the bit-error-rate (BER) performance of the proposed RS-PAC concatenated code compared to the BER performance of a PAC$(64, 32)$ code. 
For SNR values above $2.5$ dB, the error-correction performance of the RS-PAC concatenated scheme is significantly better than the one of PAC code with a coding gain of $1.3$ dB at $\text{BER} = 10^{-5}$.
For SNR values below $2.5$ dB, the PAC code performs slightly better than the RS-PAC code.

\subsection{RS-PAC Concatenated Codes with Interleaver}

\begin{figure*}[t] 
\centering
	\includegraphics [width = 0.7\textwidth]{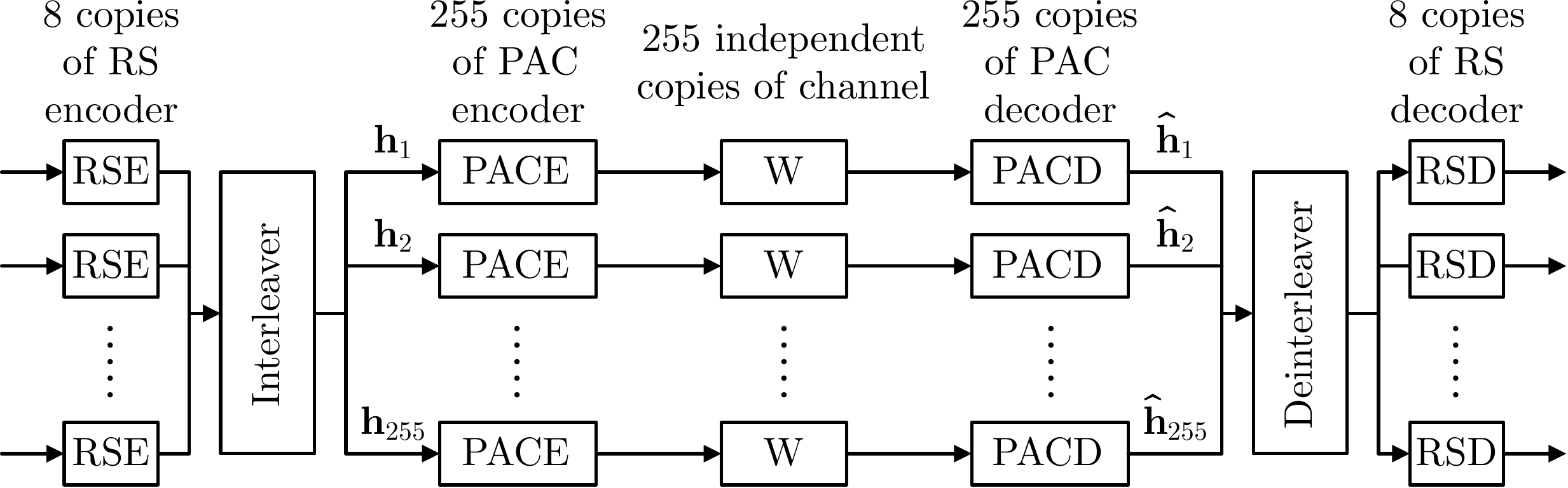}
	\caption{Block diagram of RS-PAC concatenated coding scheme interleaved to a depth of 8.} 
	\label{flowchart: RS_PAC Interleaver}
\end{figure*}

In this part, we study the effect of adopting interleaver and deinterleaver in RS-PAC concatenated code which is illustrated in Fig. \ref{flowchart: RS_PAC Interleaver}.
In concatenated codes, an interleaver and a deinterleaver are commonly used between the inner and outer codes. 
Since the deinterleaver shuffles the output of the supperchannel, possible long error bursts are distributed between multiple outer RS codes. 
In this manner, supperchannel turns into an effective random channel in which multiple outer codes can handle long error bursts.

To explain the interleaving and deinterleaving operations, we use $\mathbf{H}$ and $\mathbf{\hat{H}}$ matrices, respectively.
To interleave, the input sequences are written into the rows of the matrix $\mathbf{H}$ and the inner code reads the data from matrix $\mathbf{H}$ in column order.
To deinterleave, the output of each PAC decoder is written into the columns of matrix $\mathbf{\hat{H}}$, and the outer code reads the data from matrix $\mathbf{\hat{H}}$ in row order.

For encoding, we use eight parallel $(255, 223, 33)$ RS codes with systematic encoder as outer codes. 
We use an $8 \times 255$ matrix $\mathbf{H}$ to store the parallel RS codes' outputs. 
Each row of the matrix $\mathbf{H}$ stores the output vector of the corresponding RS encoder. 
The last 32 columns of $\mathbf{H}$ are for parity symbols. 
Consequently, $\mathbf{H}$ can be expressed as
\begin{equation*}
\begin{split}
    \mathbf{H} &= 
    ( 
    \mathbf{h}_1, \mathbf{h}_2, \cdots, \mathbf{h}_{255}
    )\\
    & =
    ( \mathbf{d}_1, \mathbf{d}_2, \cdots, \mathbf{d}_{223}, \mathbf{p}_1, \mathbf{p}_2, \cdots, \mathbf{p}_{32}
    ),
\end{split}
\end{equation*}
where the vector $\mathbf{h}_i$ is the $i$th column of matrix $\mathbf{H}$ with a length of 8 symbols (64 bits).
We use 255 parallel PAC$(128, 64)$ encoders to encode column vectors $\mathbf{h}_i$ for $i$ from 1 to 255.
The output of each PAC encoder is sent through 255 copies of the channel, and the channel outputs are decoded with the corresponding PAC decoder to obtain the estimate $\hat{\mathbf{h}}_i$ of $\mathbf{h}_i$. 
These $\hat{\mathbf{h}}_i$ vectors are stored in
\begin{equation*}
    \mathbf{\hat{H}} = 
    ( 
    \mathbf{\hat{h}}_1, \mathbf{\hat{h}}_2, \cdots, \mathbf{\hat{h}}_{255}
    ).
\end{equation*}
Finally, each row of matrix $\mathbf{\hat{H}}$ is decoded by using one of the 8 parallel RS decoders to obtain data matrix estimate
\begin{equation*}
    \hat{\mathbf{D}} = 
    (
    \hat{\mathbf{d}}_1, \hat{\mathbf{d}}_2, \cdots
    \hat{\mathbf{d}}_{223}
   ).
\end{equation*}
Each row of matrix $\hat{\mathbf{D}}$ is an estimate to the corresponding RS encoder input.
If one of the RS decoders declares a decoding failure, we use the output of the deinterleaver as the estimate of the data. 
Otherwise, we use the outputs of RS decoders.
Alternatively, we can use four or five copies of $(255, 223, 33)$ RS codes as an outer code and 255 copies of PAC$(64, 32)$ or PAC$(64, 40)$ codes as inner codes, respectively.

\begin{figure}[t] 
\centering
	\includegraphics [width = 0.95\columnwidth]{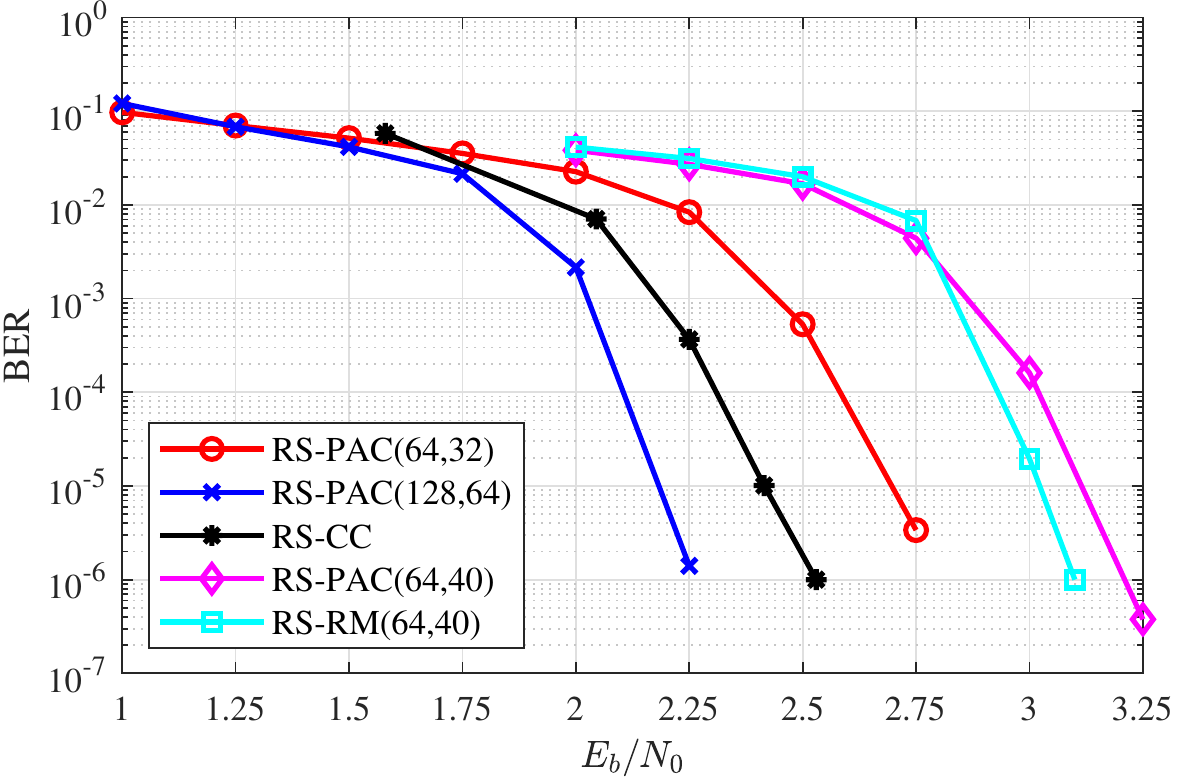}
	\caption{RS-PAC v. RS-CC and RS-RM concatenated codes with interleaver.} 
	\label{fig: BER interleaver}
\end{figure}

Fig. \ref{fig: BER interleaver} demonstrates the BER performance of the proposed RS-PAC coding scheme when using PAC$(128, 64)$ and PAC$(64, 32)$ as the inner code. We compare the performance of RS-PAC$(64, 32)$ and RS-PAC$(128, 64)$ with the NASA standard RS-CC code \cite[p.~761]{lin2001error}.
This standard uses $(255, 223, 33)$ RS code as the outer code and a rate $1/2$ 64-state CC generated by two polynomials $g_1(x) = 1+x+x^3+x^4+x^6$ and $g_2(x) = 1+x^3+x^4+x^5+x^6$ as the inner code. 
This scheme has been employed (with ideal interleaver) by NASA in some deep-space missions.
Compared to RS-CC, RS-PAC$(128, 64)$ has approximately $0.25$ dB coding gain, whereas RS-PAC$(64, 32)$ has $0.25$ coding loss at $\text{BER} = 10^{-5}$.
As the results show, in terms of error correction performance, concatenating RS codes with PAC$(128, 64)$ codes is more favorable than concatenating RS codes with CC codes or RS-PAC$(64, 32)$ codes. 
Notice that the number of parallel RS codes employed in RS-PAC$(64, 32)$ code is $4$, which is half the ones used in RS-PAC$(128, 64)$ and RS-CC.
In terms of the number of outer codes, the RS-PAC$(64, 32)$ concatenation scheme results in less complexity.

Fig. \ref{fig: BER interleaver} also compares the BER performance of concatenating RS codes and PAC$(64, 40)$ code against RS-RM$(64, 40)$ scheme reported in \cite{morelos1999constructions}.
For both of them, $5$ parallel copies of $(255, 223, 33)$ RS codes are used as outer codes, and $255$ copies of PAC$(64, 40)$ and RM$(64, 40)$ codes are used as inner codes.
Compared to RS-RM$(64, 40)$, RS-PAC$(64, 40)$ has approximately $0.1$ dB coding loss at $\text{BER} = 10^{-6}$.

\begin{figure}[t] 
\centering
	\includegraphics [width = 0.95\columnwidth]{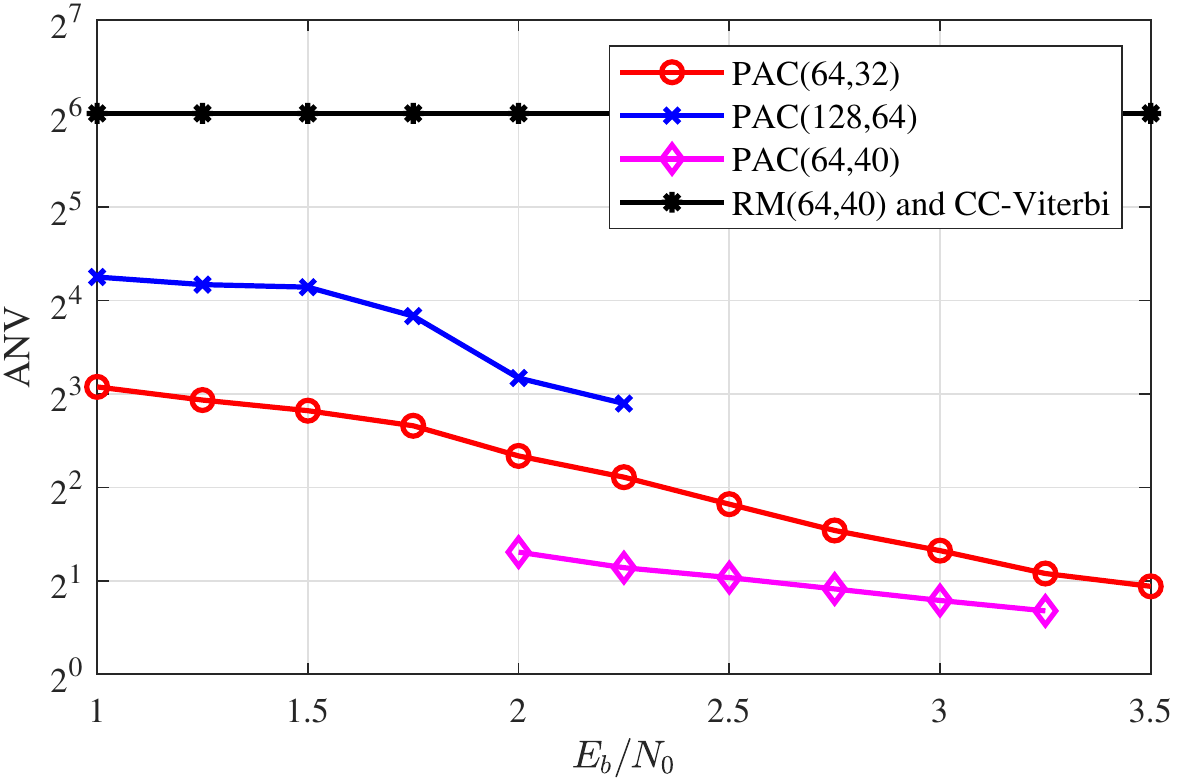}
	\caption{ANV of PAC codes v. ANV of convolutional and RM codes.} 
	\label{fig: ANV}
\end{figure}

Besides error-correction performance, the decoding complexity of a coding scheme is an important comparison factor, especially in the case of practical implementations.
Since the outer codes of concatenation schemes of Fig. \ref{fig: BER interleaver} are all RS codes of the same decoding complexity, it makes sense to compare the decoding complexity of inner codes instead of comparing the overall decoding complexity.
To measure the complexity of inner decoders, we use the notion of the average number of visits (ANV) used in \cite{moradi2021monte}, which denotes the average number of times each bit is visited by the sequential decoder during a decoding session \cite[p.~444]{wozencraft1957sequential}.
The decoding complexity of inner decoders of concatenation schemes of Fig. \ref{fig: BER interleaver} is plotted in Fig. \ref{fig: ANV}.
The Viterbi and RM decoders have fixed ANV, which depends on the number of states of their code trellises. 
The CC of RS-CC scheme and RM code of RS-RM scheme both have $64$ states, which results in a fixed ANV value of $64$.
On the other hand, the ANV of sequential decoders of RS-PAC schemes are variable and SNR dependent such that as SNR increases, the ANV decreases. 
As the results show, the ANV values of PAC$(128, 64)$ and PAC$(64, 32)$ decoders are much less than the fixed ANV of Viterbi decoding.
Although RS-PAC$(64, 32)$ has the worst BER performance compared to its counterpart concatenated schemes, its inner code has the lowest ANV.
Also, while the BER performance of RS-PAC$(64, 40)$ is slightly worse than the BER performance of RS-RM$(64, 40)$, its inner decoder complexity is significantly lower than the one of RS-RM$(64, 40)$.

\section{Conclusion}\label{sec: Conclusion}
We proposed two concatenated coding schemes that use PAC codes and RS codes as the inner and outer codes.
We evaluated the BER and complexity performance of the proposed schemes and provided comparisons with similar concatenation schemes from the literature.
Simulation results showed that concatenating PAC codes with RS codes significantly improves its error-correction performance.
Results also showed that RS-PAC codes have significantly lower decoding complexity compared to RS-RM and RS-CC codes of the same code rate while having superior error-correction performance when a proper PAC code is chosen.

\bibliographystyle{IEEEtran}

\begin{thebibliography}{10}
\providecommand{\url}[1]{#1}
\csname url@samestyle\endcsname
\providecommand{\newblock}{\relax}
\providecommand{\bibinfo}[2]{#2}
\providecommand{\BIBentrySTDinterwordspacing}{\spaceskip=0pt\relax}
\providecommand{\BIBentryALTinterwordstretchfactor}{4}
\providecommand{\BIBentryALTinterwordspacing}{\spaceskip=\fontdimen2\font plus
\BIBentryALTinterwordstretchfactor\fontdimen3\font minus
  \fontdimen4\font\relax}
\providecommand{\BIBforeignlanguage}[2]{{%
\expandafter\ifx\csname l@#1\endcsname\relax
\typeout{** WARNING: IEEEtran.bst: No hyphenation pattern has been}%
\typeout{** loaded for the language `#1'. Using the pattern for}%
\typeout{** the default language instead.}%
\else
\language=\csname l@#1\endcsname
\fi
#2}}
\providecommand{\BIBdecl}{\relax}
\BIBdecl

\bibitem{reed1960polynomial}
I.~S. Reed and G.~Solomon, ``Polynomial codes over certain finite fields,''
  \emph{Journal of the society for industrial and applied mathematics}, vol.~8,
  no.~2, pp. 300--304, 1960.

\bibitem{arikan2019sequential}
E.~Ar{\i}kan, ``From sequential decoding to channel polarization and back
  again,'' \emph{arXiv preprint arXiv:1908.09594}, 2019.

\bibitem{forney1965concatenated}
G.~D. Forney, ``Concatenated codes.'' 1965.

\bibitem{morelos1999constructions}
R.~Morelos-Zaragoza, T.~Fujiwara, T.~Kasami, and S.~Lin, ``Constructions of
  generalized concatenated codes and their trellis-based decoding complexity,''
  \emph{IEEE Transactions on Information Theory}, vol.~45, no.~2, pp. 725--731,
  1999.

\bibitem{falconer1969hybrid}
D.~Falconer, ``A hybrid coding scheme for discrete memoryless channels,''
  \emph{The Bell System Technical Journal}, vol.~48, no.~3, pp. 691--728, 1969.

\bibitem{moradi2020metric}
M.~Moradi, ``On the metric and computation of {PAC} codes,'' \emph{arXiv
  preprint arXiv:2012.05511}, 2020.

\bibitem{moradi2020PAC}
M.~Moradi, A.~Mozammel, K.~Qin, and E.~Ar{\i}kan, ``Performance and complexity
  of sequential decoding of {PAC} codes,'' \emph{arXiv preprint
  arXiv:2012.04990}, 2020.

\bibitem{moradi2021monte}
M.~Moradi and A.~Mozammel, ``A monte-carlo based construction of
  polarization-adjusted convolutional ({PAC}) codes,'' \emph{arXiv preprint
  arXiv:2106.08118}, 2021.

\bibitem{arikan2020systematic}
E.~Ar{\i}kan, ``Systematic encoding and shortening of {PAC} codes,''
  \emph{Entropy}, vol.~22, no.~11, p. 1301, 2020.

\bibitem{bose1960class}
R.~C. Bose and D.~K. Ray-Chaudhuri, ``On a class of error correcting binary
  group codes,'' \emph{Information and control}, vol.~3, no.~1, pp. 68--79,
  1960.

\bibitem{sugiyama1975method}
Y.~Sugiyama, M.~Kasahara, S.~Hirasawa, and T.~Namekawa, ``A method for solving
  key equation for decoding {G}oppa codes,'' \emph{Information and Control},
  vol.~27, no.~1, pp. 87--99, 1975.

\bibitem{berlekamp2015algebraic}
E.~R. Berlekamp, \emph{Algebraic coding theory (revised edition)}.\hskip 1em
  plus 0.5em minus 0.4em\relax World Scientific, 2015.

\bibitem{chien1964cyclic}
R.~Chien, ``Cyclic decoding procedures for {B}ose-{C}haudhuri-{H}ocquenghem
  codes,'' \emph{IEEE Transactions on information theory}, vol.~10, no.~4, pp.
  357--363, 1964.

\bibitem{forney1965decoding}
G.~Forney, ``On decoding {BCH} codes,'' \emph{IEEE Transactions on information
  theory}, vol.~11, no.~4, pp. 549--557, 1965.

\bibitem{lin2001error}
S.~Lin and D.~J. Costello, \emph{Error control coding}.\hskip 1em plus 0.5em
  minus 0.4em\relax Prentice hall, 2001, vol.~2, no.~4.

\bibitem{wozencraft1957sequential}
J.~M. Wozencraft, ``Sequential decoding for reliable communication,'' 1957.

\end{thebibliography}

\end{document}